\documentclass[aps,prl,showpacs,twocolumn,groupedaddress]{revtex4}  
\usepackage{graphicx}  
\usepackage{dcolumn}   
\usepackage{bm}        
\usepackage{amssymb}   
\usepackage{multirow}
\usepackage{epsfig}

\newcommand{\met}{\mbox{$\rlap{\kern0.15em/}E_T$}}

\newcommand{\wen}{\ensuremath{W \rightarrow e \nu}}

\newcommand{\zee}{\ensuremath{Z \rightarrow ee}}

\renewcommand{\appendix}{
  \renewcommand{\section}{
      \newpage\thispagestyle{plain}
      \secdef\Appendix\sAppendix}
  \setcounter{section}{0}
  \renewcommand{\thesection}{\Alph{section}}
}

\newcommand{\Appendix}[2][?]{
  \refstepcounter{section}
  \addcontentsline{toc}{appendix}
      {\protect\numberline{\appendixname~\thesection} #1}
  {\flushleft\large\bfseries\appendixname\ \thesection\par

   \nohyphens\centering#2\par}
  \sectionmark{#1}\vspace{\baselineskip}
}

\newcommand{\sAppendix}[1]{
  {\flushleft\large\bfseries\appendixname\par
   \nohyphens\centering#1\par}
  \vspace{\baselineskip}
}

\newcommand{\nohyphens}{\hyphenpenalty=10000\exhyphenpenalty=10000\relax}

\begin{document}

\hspace{5.2in} \mbox{\hfill Fermilab-Pub-08/249-E}

\title{Measurement of the electron charge asymmetry in $p\bar{p} \rightarrow W +X \rightarrow e\nu +X $ events at $\sqrt{s}=1.96$ TeV}
%
\author{V.M.~Abazov$^{36}$}
\author{B.~Abbott$^{75}$}
\author{M.~Abolins$^{65}$}
\author{B.S.~Acharya$^{29}$}
\author{M.~Adams$^{51}$}
\author{T.~Adams$^{49}$}
\author{E.~Aguilo$^{6}$}
\author{M.~Ahsan$^{59}$}
\author{G.D.~Alexeev$^{36}$}
\author{G.~Alkhazov$^{40}$}
% ...

\affiliation{\vspace{0.1 in}(The D\O\ Collaboration)\vspace{0.1 in}}
% ...
\affiliation{$^{6}$University of Alberta, Edmonton, Alberta, Canada,
                Simon Fraser University, Burnaby, British Columbia, Canada,
                York University, Toronto, Ontario, Canada, and
                McGill University, Montreal, Quebec, Canada}
% ...
\affiliation{$^{29}$Tata Institute of Fundamental Research, Mumbai, India}
% ...
\affiliation{$^{36}$Joint Institute for Nuclear Research, Dubna, Russia}
% ...
\affiliation{$^{40}$Petersburg Nuclear Physics Institute,
                St. Petersburg, Russia}
% ...
\affiliation{$^{49}$Florida State University, Tallahassee, Florida 32306, USA}
% ...
\affiliation{$^{51}$University of Illinois at Chicago,
                Chicago, Illinois 60607, USA}
% ...
\affiliation{$^{59}$Kansas State University, Manhattan, Kansas 66506, USA}
% ...
\affiliation{$^{65}$Michigan State University,
                East Lansing, Michigan 48824, USA}
% ...
\affiliation{$^{75}$University of Oklahoma, Norman, Oklahoma 73019, USA}
% ...
\date{July 21, 2008}

\begin{abstract}
We present a measurement of the electron charge asymmetry
in $p\bar{p} \rightarrow W+X \rightarrow e\nu+X$ events at 
a center of mass energy of 1.96 TeV using 0.75 fb$^{-1}$ of data collected
with the D0 detector at the Fermilab Tevatron Collider. The asymmetry
is measured as a function of the electron transverse momentum and pseudorapidity 
in the interval $(-3.2, 3.2)$ and is compared with expectations from next-to-leading order
calculations in perturbative quantum chromodynamics. These 
measurements will allow more accurate determinations of the proton parton distribution functions. 
\end{abstract}

\maketitle

In proton-antiproton scattering, $W^+$ ($W^-$) bosons are produced primarily by the annihilation 
of $u$ ($d$) quarks in the proton with $\bar{d}$ ($\bar{u}$) quarks in the antiproton.  Any 
difference between the $u$- and $d$-quark parton distribution functions (PDFs) 
will result in an asymmetry in the $W$ boson rapidity distribution between $W^+$ and $W^-$ boson production.  In this Letter we present a 
new measurement of this asymmetry with much larger statistical precision and over a wider kinematic range 
than previous measurements~\cite{CDF_RunI_II, D0_RunII}.  This information provides improved constraints on the PDFs, 
which should lead not only to reduced theoretical uncertainties in precision determinations of the 
$W$ boson mass, but also predictions for the Higgs boson production at the Tevatron and at future 
hadron colliders. Throughout this Letter, we use the notation ``electron" to mean
``electron and positron", unless specified otherwise.

We detect $W$ bosons via their decay $\wen$.  The boson rapidity ($y_W$)
can not be measured due to the unknown longitudinal momentum of the 
neutrino. We instead measure the electron charge 
asymmetry, which is a convolution of the $W$ boson production asymmetry and the parity violating 
asymmetry from the $W$ boson decay. Since the $V$-$A$ 
interaction is well understood, the lepton charge asymmetry retains sensitivity to the 
underlying $W$ boson asymmetry. The electron charge asymmetry ($A(\eta^e)$) is defined as:
\begin{eqnarray}
A(\eta^e) = \frac{d\sigma^+/d\eta^e - d\sigma^-/d\eta^e}{d\sigma^+/d\eta^e + d\sigma^-/d\eta^e},
\end{eqnarray}
where $\eta^e$ is the pseudorapidity of the electron \cite{d0_coordinate} and $d\sigma^{+}/d\eta^e$ 
($d\sigma^{-}/d\eta^e$) 
is the differential cross section for the electrons from $W^+$ ($W^-$) bosons as 
a function of the electron pseudorapidity. When the detection efficiencies and acceptances for 
positrons and electrons are identical, the asymmetry becomes 
the difference in the number of positron and electron 
events over the sum, and some systematic uncertainties on these
quantities do not affect $A(\eta^e)$. 

In this Letter we present results obtained from more than twice the 
integrated luminosity of previous measurements by the CDF \cite{CDF_RunI_II} 
and D0 \cite{D0_RunII} collaborations and extend the measurement for leptons with 
$|\eta^{\ell}|<3.2$, compared to $|\eta^{\ell}|<2.5$ for CDF and 
$|\eta^{\ell}|<2.0$ for the previous D0 measurement. By extending  
to higher rapidity leptons, we can provide information about the 
PDFs for a broader $x$ range ($0.002<x<1.0$ for $|y_W|<3.2$) 
at high $Q^2$ $\sim M^2_W$, 
where $Q^2$ is the momentum transfer squared, $x$ is the fraction of momentum of 
the proton carried by the parton and $M_W$ is the $W$ boson mass. \\

\indent The data sample used in this measurement was collected 
with the D0 detector \cite{d0det} at the Fermilab Tevatron
Collider using a set of inclusive single-electron triggers based only
on calorimeter information. The integrated luminosity is $750 \pm 46$ pb$^{-1}$ \cite{d0lumi}. 

The D0 detector includes a central tracking system, composed of a
silicon microstrip tracker (SMT) and a central fiber tracker (CFT), 
both located within a 2~T superconducting solenoidal magnet and
covering pseudorapidities 
of $|\eta_{D}|<3.0$ and $|\eta_{D}|<2.5$ respectively \cite{d0_coordinate}.
Three liquid argon and uranium calorimeters provide coverage out to 
$|\eta_{D}| \approx 4.2$: a central section (CC) with coverage 
of $|\eta_{D}|<1.1$ and two end calorimeters (EC) with a coverage of $1.5<|\eta_{D}|<4.2$ 
All three sections are longitudinally segmented into
electromagnetic and hadronic parts respectively. 

$W$ boson candidates are identified by one
isolated electromagnetic cluster accompanied by large 
missing transverse energy ($\met$). $\met$ is determined 
by the vector sum of the transverse components of the energy 
deposited in the calorimeter and the transverse momentum ($E_T$) of the 
electron. Electron candidates are further required to 
have shower shapes consistent with that of an electron.  
The $E_T$ of the electron and the $\met$ are 
required to be greater than $25$ GeV. Additionally, the transverse
mass $M_T$ of the electron and $\met$ is required to be greater
than $50$ GeV, where $M_T = \sqrt{2 E_T \met (1-\cos \Delta \phi)}$, 
and $\Delta\phi$ is the azimuthal angle between the electron and $\met$.

Electrons are required to fall within the fiducial region of the calorimeters,
and must be spatially matched to a reconstructed track in the central tracking system. Because of the different 
geometrical coverage of the calorimeters and the tracker, the electrons are 
divided into four different types depending on the locations of the 
electrons in the calorimeter and the associated track polar angle and the
collision vertex: CC electrons within the full coverage of the CFT, 
EC electrons within the full coverage of the CFT, EC electrons within 
the partial coverage of the CFT, and EC electrons outside the coverage 
of the CFT. Optimized choices for selection criteria are established for each type.  
A total of 491,250 events satisfy the selection, with 358,336 events 
with electrons in the CC and 132,914 events with electrons in the EC.
The charge asymmetry is measured in 24 electron pseudorapidity bins 
for $|\eta^e|<3.2$.\\ 

\indent The asymmetry measurement is sensitive to misidentification of the electron charge. 
We measure the charge misidentification rate with $Z \rightarrow ee$ events using a 
``tag-and-probe" method \cite{tag-probe} 
where a track matched to one electron tags the charge of the other. Tight conditions are
applied on the tag electron to make sure its charge is correctly determined. The result from the 
tag-and-probe method is corroborated using the fraction of same-sign events 
observed in data in the $Z$ boson pole region. The rate ranges from 0.2\% at 
$|\eta^e| \approx 0$ to 9\% at $|\eta^e|\approx 3$. The absolute uncertainty 
in the charge misidentification changes from 0.1\% to 2.6\% depending on the 
electron pseudorapidity, and is dominated by the statistics of the $Z$ boson sample. 

Sources of charge bias in the event selection are investigated 
by studying $\zee$ events. All selection efficiencies are measured for 
electrons and positrons separately, and no charge dependent biases in 
acceptance or efficiencies are found. To reduce any possible 
residual charge determination biases due to instrumental effects, 
the direction of the magnetic field in the solenoidal magnet was regularly 
reversed. Approximately 46\% of the selected $W$ bosons were collected 
with the solenoid at forward polarity, and 54\% at reverse polarity. 
The charge asymmetry is measured separately for each solenoid polarity 
and no significant differences are observed. \\

\indent Three sources of background can dilute the charge asymmetry: 
$\zee$ events where one electron is not detected by the calorimeter, 
$W \rightarrow \tau \nu \rightarrow e \nu \nu \nu$ events,
and multijet events in which one jet is misidentified as an electron and 
a large $\met$ is produced by fragmentation fluctuations or misreconstruction. The $A(\eta^e)$ 
values are corrected for the backgrounds in each bin. 

Events with electrons from $\zee$ and $W \rightarrow \tau \nu \rightarrow e \nu \nu \nu$ decays 
exhibit charge asymmetries, and these two background contributions are evaluated 
using Monte Carlo (MC) events generated with {\sc pythia}~\cite{pythia} 
and processed with a detailed detector simulation based on {\sc geant}~\cite{geant}. 
The fractions of $\zee$ and $W \rightarrow \tau \nu \rightarrow e \nu \nu \nu$ events
estimated to contribute to the candidate sample are (1.3~$\pm$~0.1)\% and (2.1~$\pm$~0.1)\%, respectively.

The background fraction from multijet events is estimated by starting from a
sample of candidate events with loose shower shape requirements and
then selecting a subset of events which satisfy the final tighter requirement. 
From $\zee$ events, and a sample of multijet
events passing the preselection but with low $\met$, we determine the
probabilities with which real and fake electrons will pass the final
shower shape requirement.  These two probabilities (verified to be charge symmetric), along with the
number of events selected in the loose and tight samples allow us to
calculate the fraction of multijet events within our final selection.
The final background contamination from multijet events is estimated
to be $(0.8 \pm 0.4)\%$.

The final charge asymmetry is corrected for 
electron energy scale and resolution, $\met$ resolution and trigger 
efficiency. The correction is estimated by comparing the asymmetry
from the generator level {\sc pythia} $\wen$ MC to the {\sc geant}-simulated 
results for each electron type. \\

\indent The electron charge asymmetry is determined separately for 
each electron pseudorapidity bin and for each of the four electron types and 
then combined. The charge misidentification and background estimations are 
performed independently for each of these measurements.
Assuming $A(-\eta^{e}) = -A(\eta^{e})$ due to CP invariance, we
fold the data to increase the available statistics and obtain a
more precise measurement of $A(\eta^e)$.  

\begin{figure}[h]
\begin{center}
\epsfig{file=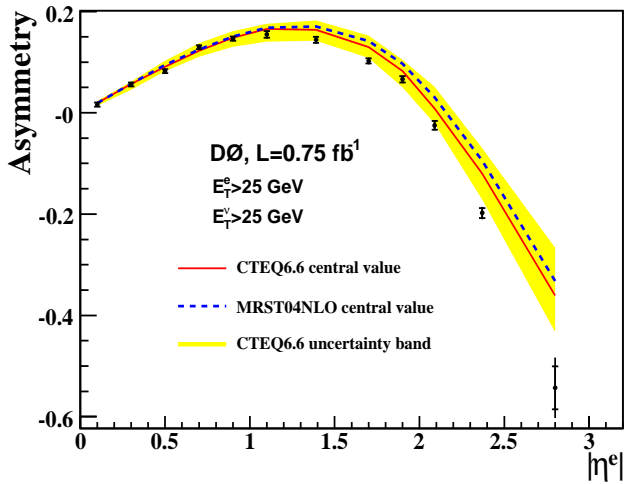, width=8.8cm, height=6.7cm}
\caption{\small (color online) The folded electron charge asymmetry distribution. 
The horizontal bars show the statistical uncertainty and the full vertical lines
show the total uncertainty on each point.
The solid line is the theoretical prediction for the asymmetry using {\sc cteq6.6} 
central PDF set. The dashed line shows the same prediction using the {\sc mrst04nlo} 
PDFs. The shaded band is the uncertainty band determined using the 44 
{\sc cteq6.6} PDF uncertainty sets. All three were determined using 
{\sc resbos} with {\sc photos}. 
}
\label{fig:wasy-unfold}
\end{center}
\end{figure}

Figure~\ref{fig:wasy-unfold} shows the folded electron charge asymmetry.
The dominant sources of systematic uncertainties originate from the
estimation of charge misidentification and multijet backgrounds.
The bin-by-bin correlations of these systematic uncertainties are negligible.
Also shown in Fig.~\ref{fig:wasy-unfold} are the theoretical predictions obtained using
the {\sc resbos} event generator~\cite{resbos} (with gluon resummation 
at low boson $p_T$ and NLO perturbative QCD calculations at high boson $p_T$) 
with {\sc photos}~\cite{photos} (for QED final state radiation). 
The PDFs used to generate these predictions are the {\sc cteq6.6} NLO PDFs~\cite{cteq} 
and {\sc mrst04nlo} PDFs~\cite{mrst}. 
Theoretical uncertainties derived from the 44 {\sc cteq6.6} PDF 
uncertainty sets are also shown. These curves are generated by 
applying a 25~GeV cut on the electron and neutrino generator-level 
transverse momenta. The asymmetric PDF uncertainty band is calculated using
the formula described in Ref.~\cite{cteq-formula}.

\begin{figure*}
\begin{center}
\epsfig{file=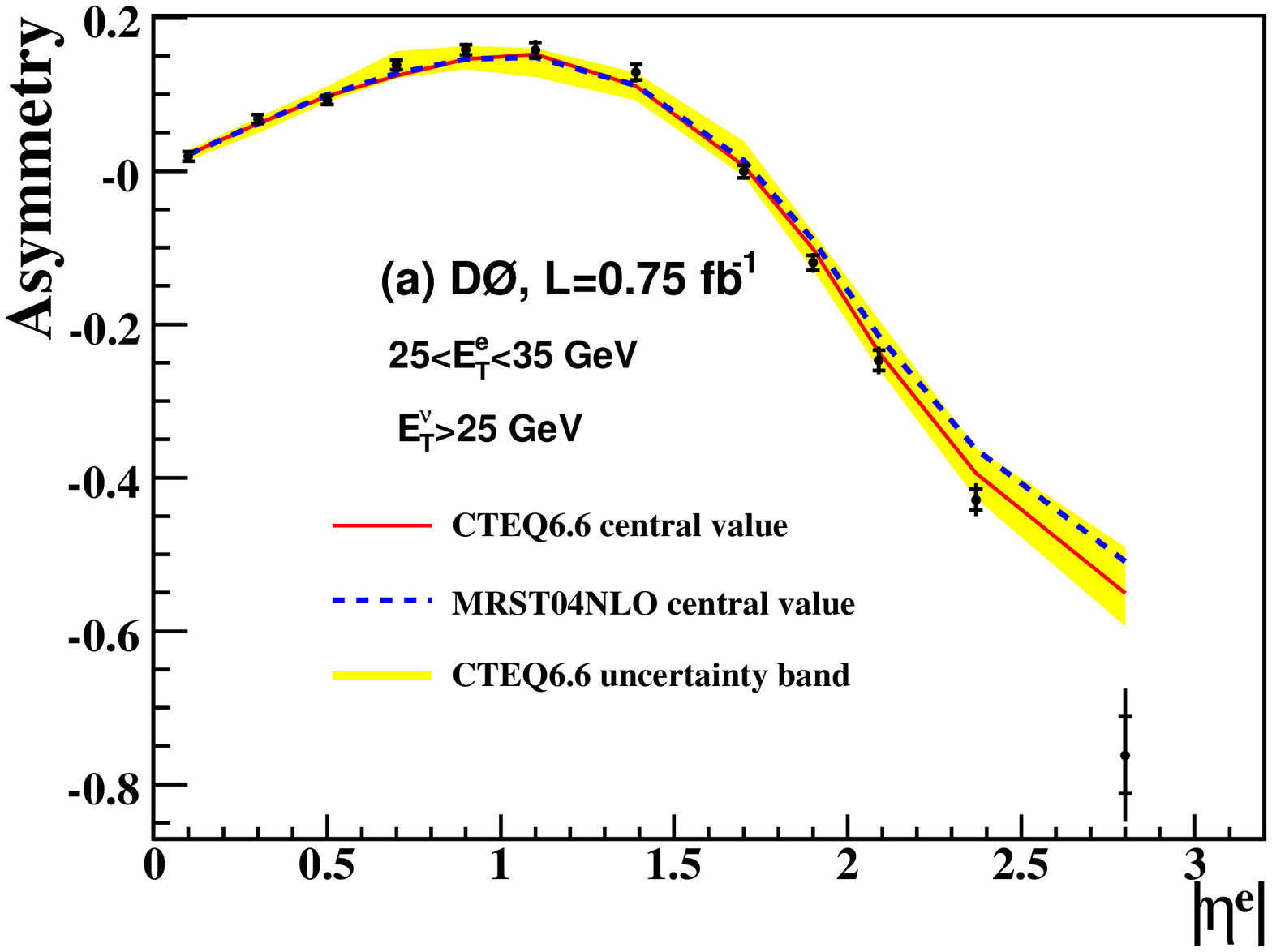, width=8.8cm, height=6.7cm}
\epsfig{file=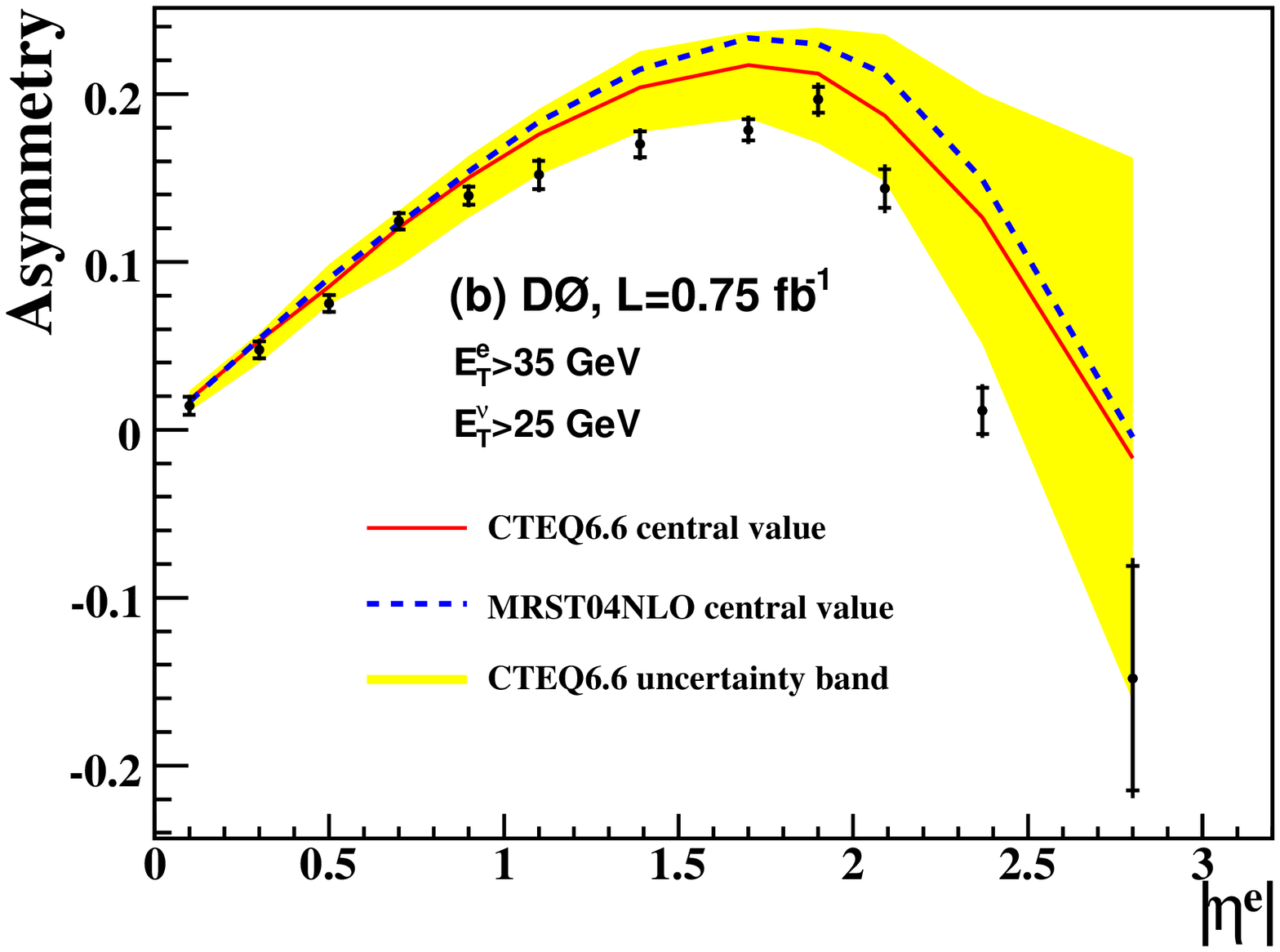, width=8.8cm, height=6.7cm}
\caption{\small (color online) The folded electron charge asymmetry distribution in two 
electron $E_T$ bins: $25<E_T<35$ GeV for (a) and $E_T>35$ GeV for (b). 
In each plot, the horizontal bars show the statistical uncertainty and the full vertical lines
show the total uncertainty on each point.
The solid line is the theoretical prediction for the asymmetry 
using {\sc cteq6.6} central PDF set. The dashed line shows the same prediction 
using the {\sc mrst04nlo} PDF set. The shaded band is the uncertainty band determined 
using the 44 {\sc cteq6.6} PDF uncertainty sets. 
All three were determined using {\sc resbos} with {\sc photos}. 
}
\label{fig:wasy-fold-pt}
\end{center}
\end{figure*}

We also measure the asymmetry in two bins of electron $E_T$: 
$25<E_T<35$ GeV and $E_T>35$ GeV. For a given $\eta^e$, the two $E_T$ regions
probe different ranges of $y_W$ and thus allow a finer probe of 
the $x$ dependence. The folded electron charge asymmetries, along with
the theoretical predictions, for the two electron $E_T$ bins are 
shown in Fig.~\ref{fig:wasy-fold-pt}.

The measured values of the asymmetry and uncertainties, together with  
the {\sc cteq6.6} predictions, for $E_T>25$~GeV and the 
two separate $E_T$ bins are listed in Table~\ref{Tab:asym_final_folded_summary}. 
The measured charge asymmetries tend to be lower than the theoretical 
predictions using both {\sc cteq6.6} central PDF set and 
{\sc mrst04nlo} PDFs for high pseudorapidity electrons. For most 
$\eta^e$ bins, the experimental uncertainties are smaller than 
the uncertainties given by the most recent {\sc cteq6.6} uncertainty sets,
demonstrating the sensitivity of our measurement. 

In summary, we have measured the charge asymmetry of electrons in
$p\bar{p} \rightarrow W+X \rightarrow e\nu+X$ using 0.75~fb$^{-1}$ of data. The electron coverage is extended
to $|\eta^e|<3.2$ and the asymmetry is measured for electron $E_T>$~25~GeV, 
as well as two separate $E_T$ bins to improve sensitivity to the PDFs. This 
measurement is the most precise electron charge asymmetry measurement to date, 
and the experimental uncertainties are smaller than 
the theoretical uncertainties across almost 
all electron pseudorapidities. Our result can be used to improve the 
precision and accuracy of next generation PDF sets, and will help to 
reduce the PDF uncertainty for high precision $M_W$ measurements and 
also improve the predictions for the Higgs boson production at the hadron
colliders.

\begin{table*}
\begin{center}
\begin{tabular}{ccccccccc}
\hline
\hline
\multirow{3}{*}{$\eta^e$ region} & \ \multirow{3}{*}{$\langle |\eta^{e}| \rangle$} \ & \multicolumn{6}{c}{$A$ ($|\eta^{e}|$)} \\ \cline{3-8}
& & \multicolumn{2}{c}{$E_T>25$ GeV}  & \multicolumn{2}{c}{$25<E_T<35$ GeV}  &  \multicolumn{2}{c}{$E_T>35$ GeV}  \\
& & {Data} & {Prediction} & {Data} & {Prediction}   & {Data} & {Prediction} \\  \hline 
0.0 $-$ 0.2 & 0.10 & \ \ \ 1.6 $\pm$ 0.4 $\pm$ 0.3 & \ \ \ $1.9^{+0.4}_{-0.5}$ & \ \ \ \ 1.9 $\pm$ 0.6 $\pm$ 0.5 & \ \ \ $2.1^{+0.5}_{-0.8}$ & \ \ \ \ 1.4 $\pm$ 0.5 $\pm$ 0.4 & \ \ $1.8^{+0.5}_{-0.7}$  \\[1.5mm]
0.2 $-$ 0.4 & 0.30 & \ \ \ 5.6 $\pm$ 0.4 $\pm$ 0.3 & \ \ \ $5.7^{+0.4}_{-1.2}$ & \ \ \ \ 6.8 $\pm$ 0.6 $\pm$ 0.5 & \ \ \ $6.2^{+0.8}_{-1.3}$ & \ \ \ \ 4.8 $\pm$ 0.5 $\pm$ 0.4 & \ \ $5.3^{+0.5}_{-1.3}$  \\[1.5mm]
0.4 $-$ 0.6 & 0.50 & \ \ \ 8.2 $\pm$ 0.4 $\pm$ 0.3 & \ \ \ $9.1^{+1.2}_{-0.9}$ & \ \ \ \ 9.3 $\pm$ 0.6 $\pm$ 0.5 & \ \ \ $9.8^{+1.2}_{-0.8}$ & \ \ \ \ 7.5 $\pm$ 0.5 $\pm$ 0.4 & \ \ $8.5^{+1.3}_{-1.1}$  \\[1.5mm]
0.6 $-$ 0.8 & 0.70 & \ \ 13.0 $\pm$ 0.4 $\pm$ 0.3 & \ \ $12.2^{+1.5}_{-1.2}$ & \ \ \ 13.8 $\pm$ 0.6 $\pm$ 0.5 & \ \ $12.4^{+3.1}_{-0.3}$ & \ \ 12.4 $\pm$ 0.5 $\pm$ 0.4 & \ $12.1^{+1.0}_{-2.3}$  \\[1.5mm]
0.8 $-$ 1.0 & 0.90 & \ \ 14.6 $\pm$ 0.4 $\pm$ 0.3 & \ \ $14.8^{+1.3}_{-1.8}$ & \ \ \ 15.8 $\pm$ 0.7 $\pm$ 0.6 & \ \ $14.6^{+1.7}_{-1.3}$ & \ \ 13.9 $\pm$ 0.5 $\pm$ 0.4 & \ $15.0^{+1.3}_{-2.4}$  \\[1.5mm]
1.0 $-$ 1.2 & 1.10 & \ \ 15.5 $\pm$ 0.6 $\pm$ 0.5 & \ \ $16.6^{+1.0}_{-2.5}$ & \ \ \ 15.8 $\pm$ 1.0 $\pm$ 0.8 & \ \ $15.2^{+0.7}_{-3.0}$ & \ \ 15.2 $\pm$ 0.8 $\pm$ 0.6 & \ $17.6^{+1.5}_{-2.4}$  \\[1.5mm]
1.2 $-$ 1.6 & 1.39 & \ \ 14.4 $\pm$ 0.6 $\pm$ 0.5 & \ \ $16.4^{+1.8}_{-2.2}$ & \ \ \ 12.9 $\pm$ 1.0 $\pm$ 0.8 & \ \ $11.1^{+1.8}_{-1.8}$ & \ \ 17.0 $\pm$ 0.8 $\pm$ 0.6 & \ $20.4^{+2.2}_{-2.6}$  \\[1.5mm]
1.6 $-$ 1.8 & 1.70 & \ \ 10.2 $\pm$ 0.5 $\pm$ 0.4 & \ \ $13.0^{+2.3}_{-2.2}$ & \ \ $-0.1$ $\pm$ 0.8 $\pm$ 0.6 & \ \ \ $0.7^{+3.2}_{-1.3}$ & \ \ 17.9 $\pm$ 0.6 $\pm$ 0.6 & \ $21.7^{+2.0}_{-3.1}$  \\[1.5mm]
1.8 $-$ 2.0 & 1.90 & \ \ \ 6.6 $\pm$ 0.6 $\pm$ 0.5 & \ \ \ $8.3^{+2.2}_{-3.3}$ & $-12.0$ $\pm$ 1.0 $\pm$ 0.8 & $-10.1^{+2.2}_{-2.7}$ & \ \ 19.7 $\pm$ 0.8 $\pm$ 0.7 & \ $21.2^{+2.7}_{-4.1}$  \\[1.5mm]
2.0 $-$ 2.2 & 2.09 & \ $-2.5$ $\pm$ 0.9 $\pm$ 0.6 & \ \ \ $0.9^{+4.3}_{-3.0}$ & $-24.7$ $\pm$ 1.3 $\pm$ 1.2 & $-23.6^{+4.1}_{-2.2}$ & \ \ 14.4 $\pm$ 1.2 $\pm$ 0.9 & \ $18.7^{+4.8}_{-3.9}$  \\[1.5mm]
2.2 $-$ 2.6 & 2.37 & $-19.8$ $\pm$ 1.0 $\pm$ 0.7 & $-12.0^{+5.1}_{-5.1}$ & $-42.9$ $\pm$ 1.4 $\pm$ 1.6 & $-39.4^{+3.2}_{-3.3}$ & \ \ \ \ 1.1 $\pm$ 1.4 $\pm$ 0.7 & \ $12.6^{+7.4}_{-7.5}$  \\[1.5mm]
2.6 $-$ 3.2 & 2.80 & $-54.3$ $\pm$ 4.2 $\pm$ 4.2 & $-36.1^{+9.4}_{-7.2}$ & $-76.2$ $\pm$ 5.0 $\pm$ 7.1 & $-55.1^{+6.0}_{-4.3}$ & $-14.8$ $\pm$ 6.7 $\pm$ 2.6 & \ $-1.7^{+17.9}_{-14.4}$  \\[1.5mm]
 \hline \hline
\end{tabular}
\caption{Folded electron charge asymmetry for data and predictions from {\sc resbos} with {\sc photos} using {\sc cteq6.6} PDFs tabulated in percent. 
$\langle|\eta^e|\rangle$ is the cross section weighted average of electron pseudorapidity in each bin from {\sc resbos} with {\sc photos}. 
For data, the first uncertainty is statistical and the second is systematic. 
For the predictions, the uncertainties are from the PDFs only.
}
\label{Tab:asym_final_folded_summary}
\end{center}
\end{table*}

%
We thank P. Nadolsky for many useful discussions about the theoretical predictions.
We thank the staffs at Fermilab and collaborating institutions, 
and acknowledge support from the 
DOE and NSF (USA);
CEA and CNRS/IN2P3 (France);
FASI, Rosatom and RFBR (Russia);
CNPq, FAPERJ, FAPESP and FUNDUNESP (Brazil);
DAE and DST (India);
Colciencias (Colombia);
CONACyT (Mexico);
KRF and KOSEF (Korea);
CONICET and UBACyT (Argentina);
FOM (The Netherlands);
STFC (United Kingdom);
MSMT and GACR (Czech Republic);
CRC Program, CFI, NSERC and WestGrid Project (Canada);
BMBF and DFG (Germany);
SFI (Ireland);
The Swedish Research Council (Sweden);
CAS and CNSF (China);
and the
Alexander von Humboldt Foundation (Germany).
%


\begin{thebibliography}{99}
\bibitem{d0_coordinate}
D0 uses a cylindrical coordinate system with the $z$ axis running
along the beam axis in the proton direction. Angles $\theta$ and $\phi$ are the polar and 
azimuthal angles, respectively. Pseudorapidity is defined as 
$\eta=-\ln [\tan(\theta/2)]$ where $\theta$ is measured with respect
to the interaction vertex. In the massless limit, $\eta$ is equivalent to 
the rapidity $y=(1/2) \ln[(E+p_z)/(E-p_z)]$.
$\eta_{D}$ is the pseudorapidity measured with respect to the 
center of the detector. Due to the distribution of the interactions within
the detector, electrons may have larger $\eta$ than $\eta_D$.
\end{thebibliography}
\end{document}